\documentclass[11pt,twoside]{article}
\usepackage{gs}
\usepackage[a4paper, top=2cm,bottom=2.5cm,left=2.5cm,right=2.5cm]{geometry}

\usepackage{amsmath,amssymb,amsfonts}
\usepackage{graphicx}
\usepackage{textcomp}

\usepackage{algpseudocode}
\usepackage{grffile}
\usepackage[font={footnotesize}]{caption}
\usepackage{subcaption}

\usepackage[backend=bibtex,minnames=3,maxnames=3,isbn=false,url=false,eprint=false,giveninits=true,sorting=none,citestyle=numeric-comp]{biblatex}
\usepackage[final]{changes}

\usepackage[title]{appendix}

\usepackage{float}
\floatstyle{ruled}
\newfloat{algorithm}{h}{loa}
\floatname{algorithm}{Algorithm}

\DeclareMathOperator{\proj}{proj}
\DeclareMathOperator{\prox}{prox}
\DeclareMathOperator*{\argmin}{argmin}

\AtBeginBibliography{\small}

\pdfoutput=1

\begin{document}

\title{\protect
Fast and memory-efficient reconstruction of sparse Poisson data in listmode with non-smooth priors with application to time-of-flight PET
}

\author[1]{Georg~Schramm}
\author[2]{Martin~Holler}

\affil[1]{Department of Imaging and Pathology, Division of Nuclear Medicine,
          KU Leuven, Belgium}

\affil[2]{Institute of Mathematics and Scientific Computing, 
          University of Graz, Austria. MH is a member of NAWI Graz (\url{https://www.nawigraz.at}) and BioTechMed Graz (\url{https://biotechmedgraz.at}).}

\maketitle

\begin{customabstract}
\\ \textbf{Objective:} 
Complete time of flight (TOF) sinograms of state-of-the-art TOF PET scanners have a large memory 
footprint.
Currently, they contain ${\sim}4{\cdot}10^9$ data bins which amount to ${\sim}17$\,GB 
in 32\,bit floating point precision.
Moreover, their size will continue to increase with advances in the 
achievable detector TOF resolution and increases in the axial field of view.
Using iterative algorithms to reconstruct such enormous TOF sinograms becomes increasingly
challenging due to the memory requirements and the computation time needed to evaluate the
forward model for every data bin.
This is especially true for more advanced optimization algorithms such as the
stochastic primal-dual hybrid gradient (SPDHG) algorithm which allows for the use of non-smooth priors
for regularization using subsets with guaranteed convergence.
SPDHG requires the storage of additional sinograms in memory, which severely limits
its application to data sets from state-of-the-art TOF PET systems using conventional
computing hardware.

\textbf{Approach:}
Motivated by the generally sparse nature of the TOF sinograms, we propose and analyze a new 
listmode (LM) extension of the SPDHG algorithm  for image reconstruction of sparse data 
following a Poisson distribution.
\added{The new algorithm is evaluated based on realistic 2D and 3D simulationsn, 
and a real dataset acquired on a state-of-the-art TOF PET/CT system.
The performance of the newly proposed LM SPDHG algorithm is compared against
the conventional sinogram SPDHG and the listmode EM-TV algorithm.} 

\textbf{Main results:}
We show that the speed of convergence of the proposed 
LM-SPDHG is equivalent the original SPDHG operating on binned data (TOF sinograms).
However, we find that for a TOF PET system with 400\,ps TOF resolution and 25\,cm axial FOV,
the proposed LM-SPDHG reduces the required memory from approximately 56\,GB to
0.7\,GB for a short dynamic frame with $10^7$ prompt coincidences and to 12.4\,GB for a long 
static acquisition with $5\cdot10^8$ prompt coincidences.

\textbf{Significance:}
In contrast to SPDHG, the reduced memory requirements of LM-SPDHG enables 
a pure GPU implementation on state-of-the-art GPUs - avoiding memory transfers
between host and GPU - which will substantially accelerate reconstruction times.
This in turn will allow the application of LM-SPDHG in routine clinical practice where short
reconstruction times are crucial.

\end{customabstract}


\section{Introduction}

A major challenge of image reconstruction in positron emission tomography (PET)
is noise suppression since the acquired emission data suffer from high levels of Poisson
noise due to limitations in acquisition time, injectable dose and scanner sensitivity.
To limit the transfer of the data noise into the image during model-based iterative
reconstruction (MBIR), different strategies exist. 
One possibility is to add a ``smoothing'' prior to the data fidelity term in the cost
function that is being optimized.
In general, we can formulate the resulting optimization problem for any imaging system where the
acquired data follow a Poisson distribution as
\begin{equation}
\argmin _{x\geq 0} \sum_{i=1}^{m} \underbrace{(Px + s)_i -  d_i \log \left( (Px + s)_i \right)}_{D_i((Px+s)_i)} + \, \beta R(Kx),
\label{eq:primal} 
\end{equation}
where $x$ is the image to be reconstructed, $P$ is the linear forward model, $d$ are the acquired
data and $s$ are additive contaminations.
$\sum_{i=1}^m D_i((Px + s)_i)$ is the negative Poisson log-likelihood, 
$i$ is the index of the data bin and $m$ is the total number of data bins.
In the specific case of time of flight (TOF) PET, $P$ is the time of flight (TOF) PET
forward model including the effects of attenuation, normalization and limited spatial resolution, 
$d$ are the acquired prompt TOF PET coincidences (the emission sinogram), 
and $s$ are the estimated random and scattered coincidences. 

$R(K\cdot)$ is a ``smoothing prior'' consisting of a generic linear operator $K$ that calculates 
local differences and a proper, convex, lower-semicontinous function $R$.
The level of regularization is controlled by the non-negative scalar factor $\beta$.
A specific example for $K$ would be the gradient operator $\nabla$, e.g. approximated by finite forward 
differences in the discretized setting.
Combining the gradient operator for $K$ with the mixed L2-L1 norm for $R$ leads to the well-known 
Total Variation (TV) prior \cite{Rudin1992}.

The TV prior, as well as many other advanced smoothing priors aiming for edge-preservation 
such as e.g. Total Generalized Variation (TGV) 
\cite{Bredies2010}, Joint T(G)V \cite{Rigie2015,Knoll2016}
Parallel Level Sets \cite{Ehrhardt2016a,Schramm2017} or directional Total Variation (DTV)
\cite{Ehrhardt2016}, require the use of non-smooth functions for $R$ as instrumental building block.
This prevents the use of simple and efficient 
purely gradient-based optimization algorithms to solve \eqref{eq:primal}.

\subsection*{PDHG and SPDHG for PET reconstruction with non-smooth priors}

Using the fact that \replaced{$z \mapsto D(z) := \sum_{i=1}^m D_i(z_i)$ and $z 
\mapsto \beta R(z)$}{$D(Px + s) = \sum_i D_i((Px + s)_i)$ and $\beta R(Kx)$} are 
convex\added{, lower-semicontinuous} functions and thus are
equal to their convex biconjugates
\replaced{$D^{**}(z) = \sup_y \langle z, y \rangle - \sum_{i=1}^{m} D_i^*(y_i)$ 
and $(\beta R)^{**}(z) = \sup_w \langle z, w \rangle - (\beta R)^*(w)$}{$D^{**}(Px + s) = \sup_y \langle Px + s, y \rangle - \sum_{i=1}^{m} D_i^*(y_i)$ 
and $(\beta R)^{**}(Kx) = \sup_w \langle Kx, w \rangle - (\beta R)^*(w)$}, respectively, 
where $D_i^*$ and $(\beta R)^*$ are the convex conjugates,
and that $(\beta R)^*(w) = \beta R^*(w / \beta)$, 
we can rewrite \eqref{eq:primal} as the saddle point problem
\begin{equation}
\argmin_{x\geq 0} \, \sup_{y,w} \,  \langle Px + s, y \rangle + \langle Kx, w \rangle - \sum_{i=1}^{m} D_i^*(y_i) - \beta R^*(w/\beta) ,
\label{eq:saddle}
\end{equation}
introducing the dual variables $y$ and $w$, and the convex dual of the Poisson log-likelihood given as
\begin{equation}
D_i^*(y_i) =
\begin{cases}
-d_i + d_i \log \left( \frac{d_i}{1-y_i} \right) & \text{if } y_i < 1 \text{ and } d_i > 0{,} \\
0 & \text{if } y_i \leq 1 \text{ and } d_i = 0\added{,} \\
\infty & \text{else} \ .
\end{cases}
\end{equation}
%
Under mild assumptions on $R$, which hold for all of \added{the} above-mentioned smoothing priors, Problem \eqref{eq:saddle} is equivalent to \eqref{eq:primal} and
can be solved even for non-smooth priors using the generic primal-dual hybrid gradient (PDHG) 
algorithm by Chambolle and Pock \cite{Chambolle2011}.
PDHG is an iterative algorithm that requires the evaluation of the complete forward and adjoint operator
in every update.
The usage of the original PDHG algorithm to solve \eqref{eq:saddle} for real-world state-of-the-art
TOF PET systems, however, usually results in extremely long computation times, 
because the evaluation of $P$ and $P^T$
for state-of-the-art TOF PET systems is computationally very demanding, and 
because several hundreds to thousands of updates are needed to obtain reasonable convergence.

To overcome this limitation, Chambolle et al. published a stochastic extension of PDHG called SPDHG 
for saddle point problems that are separable in the dual variable in 2018 \cite{Chambolle2018}.
In contrast to PDHG, SPDHG has the advantage that the complete forward and adjoint operator are
split into $n$ subsets, and that, in every update, only a random subset of the forward
and adjoint operator chosen according to a probability $p_k$ has to be evaluated.
In \cite{Ehrhardt2019}, Ehrhardt et al. applied SPDHG to 3D non-TOF PET reconstruction with TV-like
priors and showed that around 10 complete projections and back projections 
\replaced{can be}{are} sufficient 
to obtain reasonable convergence using SPDHG with 252 subsets.
Moreover, the authors also demonstrated that preconditioning further accelerates convergence.
The resulting SPDHG algorithm to solve \eqref{eq:saddle} is summarized in Algorithm \ref{alg:spdhg},
where the proximal operator for a convex function $f$ using the weighted norm 
$\| \cdot \|_M^2 = \langle M^{-1} \cdot, \cdot \rangle$ induced by a symmetric and 
positive\added{-definite} matrix $M$ is defined as
\begin{equation}
\prox_{f}^{M}(y) = \argmin_v \left( \frac{1}{2} \| v - y  \|_M^2 + f(v) \right) \ .
\end{equation}
For the convex conjugate of the negative Poisson log-likehood $D_i^*$, this proximal operator
can be calculated point-wise and is given by
\begin{equation}
\begin{split}
(\prox_{D_i^*}^{S_i}(y))_i &= \prox_{D_i^*}^{S_i}(y_i) \\ 
&= \frac{1}{2} \left(y_i + 1 - \sqrt{ (y_i-1)^2 + 4 S_i d_i} \right) \ .
\end{split}
\label{eq:proxD}
\end{equation} 
The proximal operator for $R^*$ obviously depends on the choice of $R$ but can be also efficiently 
computed using point-wise operations for many popular choices of $R$.
As mentioned in \cite{Ehrhardt2019}, Algorithm \ref{alg:spdhg} converges if we use the preconditioned
step sizes
\[ S_k = \gamma \, \text{diag}(\frac{\rho}{P_k 1} )\qquad  T_k = \gamma^{-1} \text{diag}(\frac{\rho p_k}{P^T_k 1}) \ , \]
and
\[ S_{n+1} = \gamma \, \frac{\rho}{\|K\|} \qquad T_{n+1} = \gamma^{-1} \frac{p_{n+1}\rho}{\|K\|} \ , \]
setting $T = \min_{k=1,\ldots,n+1} T_k$ pointwise, and choosing $0<\rho<1$ and $\gamma>0$.
\begin{algorithm}[t]
\begin{algorithmic}[1]
\small
\State \textbf{Initialize} $x(=0),y(=0),(w=0)$, $(S_i)_i,T,(p_i)_i$,
\State $\overline{z} = z = P^T y + K^T w$
\Repeat
	\State $x \gets \proj_{\geq 0} (x - T \overline{z})$
	\State Select $i \in \{ 1,\ldots,n+1\} $ randomly according to $(p_i)_i$
  \If{$i \leq n$}
	\State $y_i^+ \gets \prox_{D_i^*}^{S_i} ( y_i + S_i  ( P_i x + s_i))$
	\State $\delta z \gets P_i^T (y_i^+ - y_i)$
	\State $y_i \gets y_i^+$
  \Else
	\State $w^+ \gets \beta \prox_{R^*}^{S_i/\beta} ((w + S_i  K x)/\beta)$
	\State $\delta z \gets K^T (w^+ - w)$
	\State $w \gets w^+$
  \EndIf
	\State $z \gets z + \delta z$
	\State $\overline{z} \gets  z + (\delta z/p_i)$
\Until{stopping criterion fulfilled}
\State \Return{$x$}
\end{algorithmic}
\caption{SPDHG for PET reconstruction \cite{Ehrhardt2019}}
\label{alg:spdhg}
\end{algorithm}

\subsection*{Limitations of PDHG and SPDHG for PET reconstruction}

Whilst SPDHG is a big step forward for an efficient solution of \eqref{eq:saddle}
in terms of computational speed, SPDHG also comes with two main limitations.
First, as discussed in Remark 2 of \cite{Ehrhardt2019}, 
a potential drawback of SPDHG is that it requires keeping at least one more complete 
(TOF) sinogram in memory (the dual variable $y$). 
Moreover, if the proposed preconditioning is used, a second complete (TOF) sinogram
(the sequence of step sizes $(S_k)_{k=1}^n$) needs to be stored in memory.
Storing one or two extra sinograms in memory of modern computers is not a major problem for static 
single-bed non-TOF PET data, where sinogram sizes are relatively small.
However, for simultaneous multi-bed, dynamic or TOF PET data, the size of the complete data sinograms
can be become problematic.
This issue gets even more severe when aiming for a complete GPU implementation 
of Algorithm \ref{alg:spdhg} since the available memory on state-of-the-art GPUs is nowadays
much smaller compared to CPUs.

As an example, for a TOF PET scanner with 25\,cm axial FOV and a TOF resolution of ca. 400\,ps, 
a complete unmashed static TOF sinogram for one bed position 
has approximately $4.4\cdot10^9$ data bins, requiring ca. 17\,GB of memory in 32\,bit floating
point precision.
Note that with improved TOF resolution and increasing axial field of view (e.g. total body 
PET scanners), the memory required to store a complete TOF sinogram will continue to 
substantially increase in the future.

Second, PDHG and SPDHG only work with ``binned'' data.
In the case of TOF PET reconstruction, that means that the acquired raw listmode data first need to be
binned into TOF sinograms and that $P$ and $P^T$ need to be evaluated using
sinogram projectors.
For most acquisitions with modern TOF PET scanners, this is inefficient both in terms of memory and
computational time since the TOF emission data are extremely sparse.
In contrast, storage and processing of the data in listmode format (event by event) is usually 
more efficient.

\subsection*{Sparsity of TOF PET data}

Compared to non-TOF PET emission sinograms, TOF PET emission sinograms of most acquisitions
with state-of-the-art TOF PET scanners are extremely sparse.
This is because every geometrical line of response (LOR) has to be subdivided into
several small TOF bins.
To achieve sufficient sampling of the TOF information, the number of TOF bins has to be
inversely proportional to the TOF resolution of the scanner. 
Consequently, for a fixed number of acquired prompt coincidences, the sparsity of the
sinogram is proportional to the TOF resolution.

\begin{figure}
  \centering
    \includegraphics[width=0.7\columnwidth]{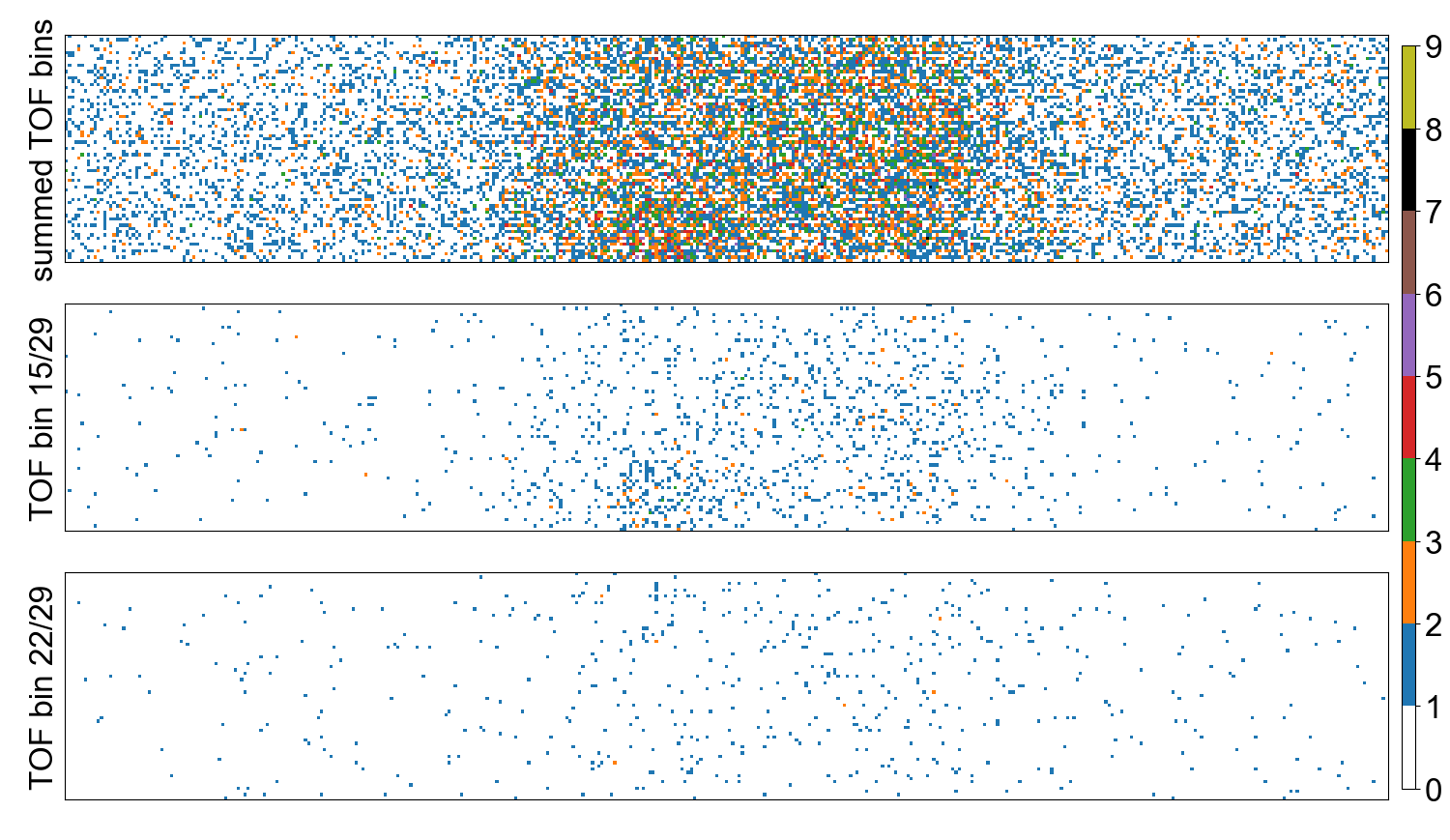}
  \caption{Representative slices through a single view of an emission sinogram of a 
  80s [\textsuperscript{18}F]FDG acquisition of a liver bed position. 
  The scan was acquired 1\,h post injection with an injected dose of 323\,MBq on
  a GE DMI PET/CT with a TOF resolution of 400\,ps (29 TOF bins). 
  The horizontal and vertical axis represent the radial and axial direction (direct planes only), 
  respectively. 
  (top) sum over all TOF bins. (middle) central TOF bin 15/29 with 94\% empty bins. 
  (bottom) TOF bin 22/29 with 97\% empty bins.}

  \label{fig:sparsity}
\end{figure}

As an example, for a typical 80\,s acquisition of a liver bed position in an FDG scan with 
an injected dose of 323\,MBq acquired 60\,min p.i. on a state-of-the-art TOF PET/CT scanner with
20\,cm axial FOV, more than 94\% of the data (TOF sinogram) bins are empty.
An example that demonstrates this extreme sparsity of TOF sinograms is shown in Fig.~\ref{fig:sparsity}.
For shorter frames, as present, e.g., in the early phase of dynamic scans or when respiratory
or cardiac gating is used, the fraction of empty bins can even higher.

Note that we expect that the sparsity of the TOF PET emission data of future PET systems will increase 
faster than linear compared to the improvement of the TOF resolution.
This is because with better TOF resolution, every detected event carries more information such
that fewer detected events are needed to reconstruct images with the similar 
variance \cite{Tomitani1981}.
Moreover, the method presented in this work is applicable not only to TOF PET image reconstruction,
but to all imaging reconstrution problems with sparse data following a Poisson distribution
such as (low count) SPECT or photon counting CT.

\subsection*{Contributions and Aim}

To improve the efficiency of SPDHG in terms of required memory and computation time when reconstructing 
sparse TOF PET data, we propose and analyze a listmode extension of the SPDHG algorithm  called
LM-SPDHG that allows event-by-event processing using dedicated listmode forward and back projectors.
We first derive LM-SPDHG from SPDHG and show that the convergence of LM-SPDHG is as 
fast as the convergence of SPDHG based on dedicated numeric examples in 2D.
Moreover, we analyze the memory requirements for LM-SPDHG compared to SPDHG for typical scans 
acquired on state-of-the-art TOF PET scanners.

We emphasize that the focus of this work is not on finding the
optimal (non-smooth) prior or prior strength for a given clinical task in PET imaging.
Instead, we aim to provide a framework enabling fast and memory efficient reconstruction of 
sparse TOF PET listmode data which will finally facilitate future research on the use of
(non-smooth) priors in PET reconstruction.

\section{Theory and Algorithms}

Before deriving LM-SPDHG, the next subsection first of all shows how to reduce the memory 
requirements of SPDHG when reconstructing sparse TOF sinograms. 

\subsection*{Memory-efficient SPDHG for sparse TOF sinograms}

As shown in \cite{Schramm2021}, the memory requirements for SPDHG can be substantially reduced by 
choosing a better initialization of the dual variable $y$.

From Eq.~(\ref{eq:proxD}) we can observe that for data bins $i$ where $d_i = 0$ 
(bins in the TOF emission sinogram with zero counts), 
$\prox_{D_i^*}(a_i) = 1$ for $a_i \geq 1$ and $\prox_{D_i^*}(a_i) = a_i$ 
otherwise. 
Moreover, provided that $y_i \geq 1$, we see that $ a_i = y_i + S_i (P_i x + s_i) \geq 1$ 
since all other quantities are non-negative. 
Consequently, if we initialize all bins of $y$ where the data sinogram $d$ equals zero with one, 
these bins of $y$ remain $1$ during all iterations. 
This in turn means that these bins do not contribute to the update of $\delta z$, $z$, $\bar{z}$, 
and $x$, since only the difference between
$y$ and $y^+$ is backprojected in line 8 of Algorithm~\ref{alg:spdhg} meaning that all
bins without data do not have to be kept in memory during the iteration loop (lines 3 until 17). 
The only place where the empty data bins contribute is the initialization of $z$ and $\bar{z}$
in line 2.
However, this single back projection can be split into smaller chunks to also reduce
the required memory of this step.

While at a first glance, the initialization of $y$ proposed above might seem artificial, in fact it 
directly corresponds to choosing the optimal value for those $y_i$ where $d_i=0$. 
To see this, note that an optimal solution $(\hat{x},\hat{y},\hat{w})$ of \eqref{eq:saddle} in 
particular needs to satisfy the optimality condition
\begin{equation}
P \hat{x} \in  \partial D^*(\hat{y}),
\end{equation}
where $\partial D^*$ is the subdifferential of $D^*$. Since this is equivalent to $\hat{y} \in  \partial D(P \hat{x})$, and since $D$ (the negative Poisson log likelihood) is differentiable this leads 
to the condition
\begin{equation}
\hat{y} = 1 - \frac{d}{P\hat{x} + s} \ ,
\label{eq:yinit}
\end{equation}
where the division is to be understood point-wise. 
With this we see that the proposed initialization directly corresponds to choosing the optimal value 
for $y_i $ where $d_i=0$, which is known explicitly in this case.

As argued above, this improved initialization of $y$ naturally reduces the 
memory requirements of SPDHG and also improves the speed of convergence when a
``warm start'' for $x^0$ is chosen. 
The latter can, e.g., be achieved by applying one iteration
and a reasonable amount of subsets of the EM-TV algorithm \cite{Sawatzky2008, Burger2008}.
Since in EM-TV every update is split into a classical EM step followed by a weighted denoising
optimization problem in image space, EM-TV can be used in listmode as well by simply
modifying the EM step.

\subsection*{Listmode SPDHG}

As indicated by the name, emission data in listmode format are a chronological list $N$ of detected 
events $e \in N$, where each event is characterized by a small set of (integer) numbers 
(e.g. the number of the two detectors, the discretized TOF difference, and a time stamp).
To process the listmode data during reconstruction without binning it into a sinogram,
we introduce the listmode forward operator $P^{LM}_N$ mapping the image data $x$ to a 
data-vector of dimension $|N|$ via 
\begin{equation}
(P^{LM}_N x)_e  = (Px)_{i_e} , \text{ for each }e \in N,
\label{eq:lmop}
\end{equation}
where $i_e$ is the sinogram bin in which event $e$ was detected. 
\added{Also, we denote by $s^{LM}$ with $s^{LM}_e = s_{i_e}$ the listmode-based scatter 
and random estimate.}

Re-writing the gradient of the negative Poisson log likelihood using listmode data and the
listmode operators is straightforward, such that any gradient-based PET reconstruction algorithm 
can be easily adapted to listmode data.

For a Bayesian approach with prior $R(K \cdot)$ as in \eqref{eq:primal} this is less immediate, 
but we can show that indeed \eqref{eq:primal} can be equivalently re-written \replaced{into}{to} 
a minimization problem involving only the listmode forward operator that is of the same form as 
\eqref{eq:primal}\added{ as shown in appendix \ref{appendix:lm}}. 
This allows us to extend the SPDHG algorithm for listmode data and yields the algorithm as shown 
in Algorithm~\ref{alg:lmspdhg}.

\begin{algorithm}[t]
\begin{algorithmic}[1]
\small
\State \textbf{Input} event list $N$, contamination list $s_N$
\State \textbf{Calculate} event counts $\mu_e$ for each e in $N$ (see text)
\State \textbf{Initialize} $x,w,(S_i)_i,T,(p_i)_i$
\State \textbf{Initialize} list $y_{N} = 1 - (\mu_N /(P^{LM}_{N} x + s_{N}))$ 
\State \textbf{Preprocessing} $\overline{z} = z = {P^T} 1 - {P^{LM}}^T (y_N-1)/\mu_N + K^T w$ 
\State \textbf{Split} lists $N$, $s_N$ and $y_N$ into $n$ sublists $N_i$, $y_{N_i}$ and $s_{N_i}$
\Repeat
	\State $x \gets \proj_{\geq 0} (x - T \overline{z})$
	\State Select $i \in \{1,\ldots,n+1\}$ randomly according to $(p_i)_i$
  \If{$i \leq n$}
	  \State $y_{N_i}^+ \gets \prox_{D^*}^{S_i} \left( y_{N_i} + S_i \left(P^{LM}_{N_i} x + \added{s^{LM}_{N_i}} \right) \right)$
	  \State $\delta z \gets {P^{LM}_{N_i}}^T \left(\frac{y_{N_i}^+ - y_{N_i}}{\mu_{N_i}}\right)$
	  \State $y_{N_i} \gets y_{N_i}^+$
  \Else
	  \State $w^+ \gets \beta \prox_{R^*}^{S_i/\beta} ((w + S_i  K x)/\beta)$
	  \State $\delta z \gets K^T \left(w^+ - w\right)$
	  \State $w \gets w^+$
  \EndIf
	\State $z \gets z + \delta z$
	\State $\overline{z} \gets  z + (\delta z/p_i)$
\Until{stopping criterion fulfilled}
\State \Return{$x$}
\end{algorithmic}
\caption{LM-SPDHG for PET reconstruction}
\label{alg:lmspdhg}
\end{algorithm}

In contrast to the original SPDHG using binned data (sinograms), the forward and adjoint
PET operators (lines 11 and 12) have been replaced by their listmode equivalents as in \eqref{eq:lmop}.
Moreover, the dual variable for the data fidelity is replaced by the list $y_N$, which has the
same length as the measured event list $N$.
If an event with a fixed sinogram bin $i_e$ occurs more than once in the event list
$N$, it is also forward and back-projected multiple times in steps 11 and 12.
To compensate for this fact, the algorithm divides by the event count\footnote{\added{the event count $\mu_e$ represents the number of times the event $e$ 
occurs in the event list $N$. For TOF PET, a data bin is characterized by the combination of geometrical LOR and the TOF bin along the LOR (a bin in a TOF sinogram).}}
$\mu_e$ before back projection in line 12.
Note that (i) as shown in Fig.~\ref{fig:sparsity} in most standard acquisitions the event count
of most events is 1 and that (ii) calculating the event count $\mu_e$ which is a prerequisit
creates a small pre-processing overhead (step 2).
However, when implemented on a modern GPU this overhead is small compared to the computation
time needed to calculate all iterations.
Moreover, it is in the similar order of the time needed to unlist the native listmode data into
a sinogram which is a prerequisit and pre-processing overhead for SPDHG.

Another difference of LM-SPDHG compared to SPDHG is the fact that we split the data into $n$ subsets by
assigning every $n$-th event of the complete event list $N$ to the $n$-th sub list - 
as commonly done when defining subsets in listmode OSEM.
In that way, we can think of the subset listmode forward operator $P^{LM}_{N_i}$
as the full forward operator $P$ with a sensitivity reduced by a factor of $n$.
Accordingly, we set the step sizes associated with the subset listmode PET operators to
\begin{equation}
S_k = \gamma \, \text{diag}(\frac{\rho}{P^{LM}_{N_k} 1} )\qquad  T_k = \gamma^{-1} \text{diag}(\frac{\rho p_k}{P^T 1/n}) \ . 
\label{eq:lm_stepsizes}
\end{equation}
At a first glance, the initialization of $z$ and $\bar{z}$ in step 5 of Algorithm~\ref{alg:lmspdhg} 
might look odd.
However, the first part of the expression is equivalent to 
applying the adjoint sinogram operator $P^T$ to a sinogram initialized according to \eqref{eq:yinit}.
Note that step 5 still requires to calculate a sinogram back projection of a unity sinogram ($P^T 1$)
which we would like to avoid in listmode data processing.
Unfortunately, avoiding this single sinogram backprojection is not possible.
However, note that this ``sensitivity image`` is also needed in the calculation of the step sizes
$T_k$ and also in the listmode EM-TV algorithm that we use to initialize $x$.
Hence, we recommend pre-computing the sensitivity image ($P^T 1$) and storing it in memory.

In summary, compared to SPDHG, LM-SPDHG as shown in Algorithm~\ref{alg:lmspdhg} has the advantage
that (i) only lists ($N$, $y_{N}$, $s_N$, $\mu_N$) instead of complete sinograms 
($y$ and $d$) have to be 
stored in memory and (ii) all projections and back projection\added{s} can be performed 
using listmode projectors.
This means that for all acquisitions where the number of detected events is substantially smaller than
the number of data bins, the required memory and computation time is reduced.
The latter depends on the actual implementation of the sinogram and listmode projectors
and on the computational hardware.
According to our experience using a state-of-the art GPU implementation of Joseph projectors 
\cite{Joseph1982} for a TOF PET scanner with 400\,ps TOF resolution and 25\,cm axial field of view,
a forward and back projection is faster in listmode if approximately less than 3e8 events 
have to be processed.
For 7e7 and 1e7 counts, the projections are approximately faster by a factor of 3 and 5,
respectively\footnote{The reported computational times for the projections include the time
needed to transfer the image and projection data to and from host to the GPU and thus
correspond to a hybrid CPU/GPU computational model.}.
Note that the difference in the computation time of the projection for sinogram and listmode 
depends on the specific implementation - especially on the optimization of the memory access - 
and the computational hardware.
A detailed comparison of the required memory for SPDHG and LM-SPDHG for different typical PET
acquisitions is listed in Table~\ref{tab:mem}.

\begin{table}
\begin{center}
\footnotesize
\begin{tabular}{ c c r r r}
         &                                    &            & counts      &          \\ 
         &                                    & 5e8        & 7e7         & 1e7      \\ \hline
         & 9 images                           &   0.45\,GB &   0.45\,GB  & 0.45\,GB \\
SPDHG    & 1 uint, 3 float sino.              &  55.6\,GB  &  55.6\,GB   & 55.6\,GB \\ \hline
LM-      & 9 images                           &   0.45\,GB &   0.45\,GB  & 0.45\,GB \\
SPDHG    & event list, 1 uint, 3 float lists  &  12.0\,GB  &   1.7\,GB   & 0.3\,GB  \\ \hline
         & 9 images                           &   0.45\,GB &   0.45\,GB  & 0.45\,GB \\
\added{EM-TV}    & event list, 2 float list   &  9.0\,GB   &   1.3\,GB   & 0.2\,GB
\end{tabular}
\end{center}
\caption{Estimation of required memory for SPDHG and LM-SPDHG assuming an image size of (300,300,125)
         and a TOF sinogram with 357 radial elemnts, 224 views, 1981 direct and oblique planes, 
         and 27 TOF bins corresponding to a TOF PET scanner with 400\,ps TOF
         resolution and 25\,cm axial FOV for three counts levels. 
         5e8 counts approximately correspond to a high count 20\,min late static FDG brain scan, 
         7e7 counts to an 80\,s static FDG body bed position 1\,h p.i., 
         and 1e7 counts to a short early frame in a dynamic acquisition.
         In this estimation, we assume that every TOF PET listmode even can be encoded using 10 bytes.}
\label{tab:mem}
\end{table}


\section{Numerical Experiments}

Recalling the fact that for most common TOF PET acquisitions, LM-SPDHG requires less memory and 
is also faster, one would probably prefer LM-SPDHG over SPDHG if the speed of convergence is the same.
In the following, we describe a set of numerical experiments that we conducted to show that this
is indeed the case.
In the absence of an analytical solution to the optimization problems \eqref{eq:primal}
and \eqref{eq:saddle}, we analyzed the convergence of LM-SPDHG and SPDHG with respect to a 
reference solution $x^*$ as done in \cite{Ehrhardt2019}.
Convergence was monitored by tracking the relative cost function
\begin{equation}
c_\text{rel}(x) = (c(x) - c(x^*)) / (c(x^0) - c(x^*)) \ ,
\end{equation}
where $c(x)$ is the cost function to be optimized in \eqref{eq:primal} and $x^0$ is the initialization
used for $x$.
Moreover, convergence was also montiored in image space by tracking the peak signal-to-noise ratio
with respect to $x^*$
\begin{equation}
\text{PSNR}(x) = 20\,\log_{10} \left( \|x^*\|_\infty/\sqrt{\text{MSE}(x,x^*)} \right) \ ,
\end{equation}
where MSE is the mean squared error.
The reference solution was obtained by running the deterministic PDHG (SPDHG without subsets)
for 20000 iterations.
Since running PDHG with 20000 iterations using realistic 3D TOF PET data takes a very long 
time (approx. 250\,h), all numerical convergence experiments were performed using simulated 2D TOF PET data.
A 2D software brain phantom with a gray to white matter contrast of 4:1 was created
based on the brainweb phantom \cite{Collins1998} and used to generate simulated 2D TOF data 
including the effects of limited spatial resolution, attenuation and a flat contamination mimicking 
random and scattered coincidences with a contamination fraction of 42\%.
The geometry and TOF resolution of the simulated 2D PET system was chosen to 
mimic one direct plane of a state-of-the art TOF PET scanner with a ring diameter of 650\,mm
and a 400\,ps TOF resolution
(sinogram dimension: 357 radial bins, 224 projection angles, 27 TOF bins).
Noisy simulated prompt emission TOF sinograms and corresponding listmode data were generated
for two different count levels (5e5 and 5e6 prompt counts).

\begin{algorithm}[t]
\begin{algorithmic}[1]
\small
\State \textbf{Input} event list $N$, contamination list $s_N$
\State \textbf{Split} lists $N$ and $s_N$ into $n$ sublists $N_i$, and $s_{N_i}$
\State \textbf{Pre-compute} sensitivity image $g = P^T 1$
\Repeat
	\State Select subset $i$
	\State $z \gets \dfrac{x\,n}{g} {P_{N_i}^{LM}}^T \dfrac{1}{P_{N_i}^{LM} + s_{N_i}}$ (subset EM step)
  \State $w \gets g / (\beta x)$
  \State $x^+ \gets \argmin_{u\geq 0} \sum_j \dfrac{w_j}{2} \left(u_j -z_j \right)^2 + R(Ku)$
\Until{stopping criterion fulfilled}
\State \Return{$x$}
\end{algorithmic}
\caption{listmode EM-TV for PET reconstruction \cite{Sawatzky2008, Burger2008}}
\label{alg:emtv}
\end{algorithm}

Unless stated otherwise, we always applied the EM-TV algorithm, summarized in Algorithm~\ref{alg:emtv}, 
using 1 iteration and 28 subsets to initialize $x^0$ and $y^0$ according to \eqref{eq:yinit}.
To solve the weighted denoising problem in step 8 of the EM-TV algorithm, we applied the
accelerated PDHG algorithm \cite{Chambolle2011} using 20 iterations.

In all SPDHG and LM-SPDHG reconstructios, the step size ratio $\gamma$ was set 
to $3 / \|x^0\|_\infty$ and $\rho$ was set to 0.999.
When splitting the data into $n$ subsets, the vector of probabilities determining 
whether an update with respect to a subset of the data or with respect to the prior is 
done was set to 
\begin{equation}
p_k = 
  \begin{cases}
  \frac{1}{2n} \ &\text{if } k \leq n \\
  \frac{1}{2}  \ &\text{else} \ ,
  \end{cases}
\end{equation}
such that on average an update with respect to the prior was done on every second update
as suggested in \cite{Ehrhardt2019}. 
In this work, we use the term iteration for $2n$ updates such that on average in every iteration
the complete data are forward and backprojected once and $n$ updates with respect to the
prior are performed.

As benchmark examples for non-smooth priors, we consider two ``Total Variation like'' priors
in this work. 
These priors have the form
\begin{equation}
  R(Kx) = \|K x\|_{2,1} = \sum_i \sqrt{\sum_j (Kx)_{ij}^2} \ ,
\end{equation}
where $\|K x \|_{2,1}$ is the sum over all entries of the pointwise Euclidean norm of $K x$
(mixed L2-L1 norm).
The proximal operator for the convex dual of this prior is given by
\begin{equation}
(\prox_{R^*}^S(w) )_i = \frac{w_i}{\max(1,|w_i|)} \ .
\end{equation}
For the linear operator $K$, we first used the finite forward difference operator resulting
in the classical Total Variation (TV) prior \cite{Rudin1992}.
Moreover, we also implemented the Directional Total Variation (DTV) prior incorporating
structural information by only considering the component of the finite difference vector in every voxel
that is perpendicular to the spatial gradient of a prior image in that voxel \cite{Ehrhardt2016}.
In our convergence experiments using DTV, we assumed perfect structural prior information
meaning that the prior image used for DTV, was the image itself with a flipped contrast.
Note that in real acquisitions the quality of the available structural prior information is
of course inferior. 
However, we do not expect that this fact strongly affects the convergence properties of 
SPDHG and LM-SPDHG. 

To demonstrate that LM-SPDHG also works for realistic 3D TOF PET listmode data,
we \deleted{finally} simulated and reconstructed 3D TOF PET data based on the 3D 
XCAT \cite{Segars2010} phantom  and a state-of-the-art
TOF PET scanner with 20\,cm axial field of view, a ring diameter of 650\,mm and a 400\,ps TOF 
resolution (sinogram dimension: 357 radial bins, 224 projection angles, 1296 direct and  
oblique planes, 27 TOF bins).
As before, he data simulation included the effects of attenuation, finite resolution and a
flat contamination with a contamination fraction of 42\%.
In total, 7e7 prompt counts were simulated corresponding approximately to a standard 
80\,s 1\,h p.i. FDG liver bed position acquisition.
Simulated listmode data were reconstructed using LM-SPDHG with \replaced{50}{100} 
iterations and 224 subsets. 

\added{Last but not least, we also reconstructed a real TOF PET data set of the NEMA image 
quality phantom acquired on a GE Discovery MI 4 ring TOF PET/CT \cite{Hsu2017} 
(20\,cm axial FOV, sinogram dimension: 357 radial bins, 
272 projection angles, 1261 direct and  oblique planes, 29 TOF bins, 400\,ps TOF resolution).
The activity contrast between the sphere inserts and the background was 10:1 and
the acquisition contained 4.7e7 prompt (2.2e7 true) coincidences.
Reconstructions using a TV prior with $\beta = 6$ were performed with EM-TV using 100 
iterations with 1 and 28 subsets, and with LM-SPDHG using 100 iterations and 224 subsets.
We also performed a reconstruction of the same data set with LM-PDHG using 20000 iterations and 
1 subset.
This reconstruction was used as a reference to calculate the relative cost and PSNR.}

\added{All reconstruction algorithms used in this work were implemented in python3 using the 
open-source ``parallelproj'' CUDA projector libraries available at 
\url{https://github.com/gschramm/parallelproj}.
An Nvidia V100 GPU with 16\,GB RAM was used to perform all reconstructions.
Upon publication of this article, an open-source implementation of the LM-SPDHG algorithm will be
made availalbe in the associated ``pyparallelproj'' python package.
Preprocessing of the NEMA image quality phantom data set was performed using GE's duetto PET 
reconstruction toolbox v2.17.}

\subsection*{SPDHG using a warm start vs a cold start}

\begin{figure}
  \centering
    \includegraphics[width=0.99\columnwidth]{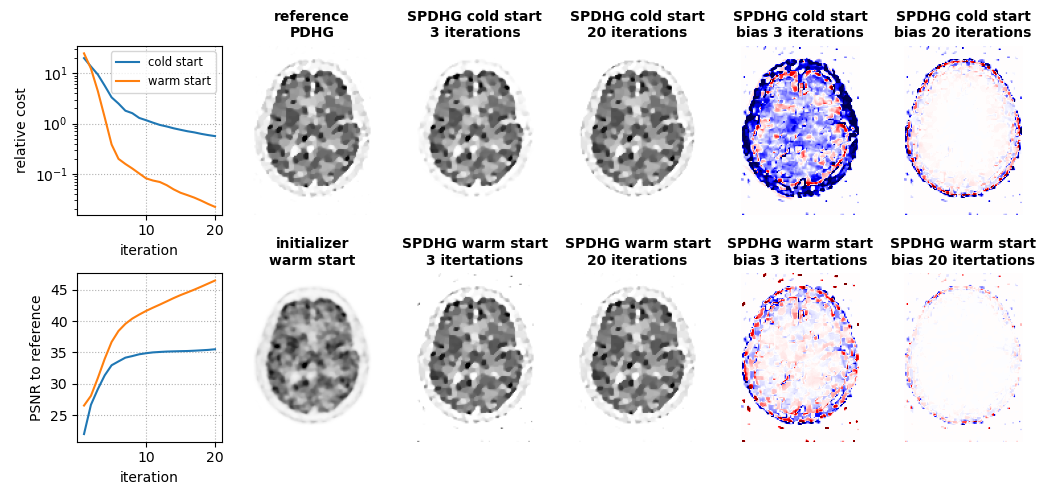}
  \caption{Comparison of convergence of sinogram SPDHG using a cold and warm start
           for 3e5 counts and a TV prior with $\beta = 0.03$ using 224 subsets.
           For the warm start, $x^0$ was taken from 1 EM-TV iteration with 28 subsets
           and $y^0$ was calculated according to \eqref{eq:yinit}.
           Left column: relative cost and PSNR with respect to the reference solution as
           a function of the iterations.
           Second column: reference PDHG reconstruction using 20000 iterations (top) and
           initializer $x^0$ used in the warm start.
           Third column: Reconstructions after 3 iterations with cold start (top) and warm
           start (bottom).
           Fourth column: Reconstructions after 20 iterations.
           \added{Fifth and sixth column: Absolute differences of the SPDHG reconstructions with
           respect to the reference PDHG reconstruction.
           The range of the color map for the difference images is $\pm$10\% of the
           maximum of the reference reconstruction.}
           Note that in the calculation of the relative cost the same $x^0$ was used
           and that in both cases the same step size ratio $\gamma$ was used.
           Moreover, in the very early iterations the relative cost of SPDHG is greater 1
           meaning that the cost is worse compared the the initialization.
           This is due to the fact that (S)PDHG in general is a non-monotonic algorithm.
          }
  \label{fig:warm_start}
\end{figure}

Before investigating the convergence behavior of LM-SPDHG, we first tested whether 
a warm start could already help to lead to faster convergence of SPDHG using sinograms.
Figure~\ref{fig:warm_start} shows the results of SPDHG reconstructions with a cold 
($x^0 = 0$ and $y^0 = 0$) and warm start as described above for a data set with 3e5 true counts
and a TV prior with $\beta = 0.03$.
It can be seen that SPDHG with the warm start performs better in terms of relative cost
and PSNR with respect to the reference reconstruction.
A similar trend was observed for higher count levels, with different $\beta$ values and using
the DTV prior indicating that as expected the warm start helps to accelerate convergence
in the early iterations.

\subsection*{Convergence of LM-SPDHG compared to SPDHG}

\begin{figure*}
  \centering
  \begin{subfigure}[]{1.0\textwidth}
    \centering
    \includegraphics[width=0.8\textwidth]{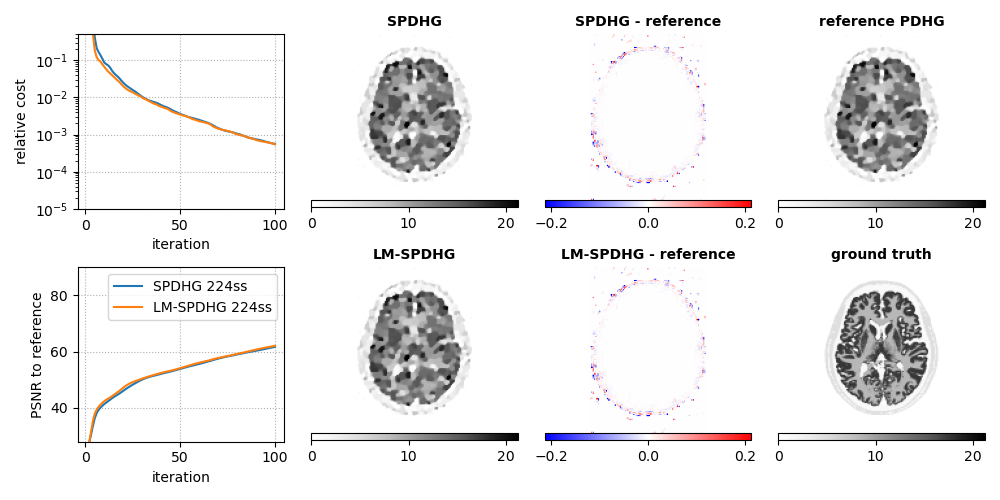}
    \caption{3e5 true (5e5 prompt) counts, TV prior, $\beta = 0.03$}
  \end{subfigure}
  \vfill
  \begin{subfigure}[]{1.0\textwidth}
    \centering
    \includegraphics[width=0.8\textwidth]{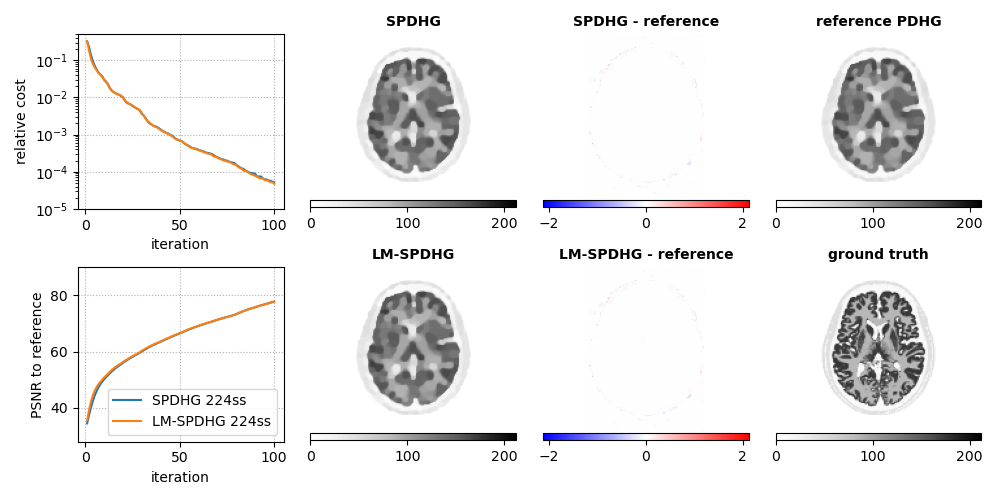}
    \caption{3e6 true (5e6 prompt) counts, TV prior, $\beta = 0.03$}
  \end{subfigure}
  \vfill
  \begin{subfigure}[]{1.0\textwidth}
    \centering
    \includegraphics[width=0.8\textwidth]{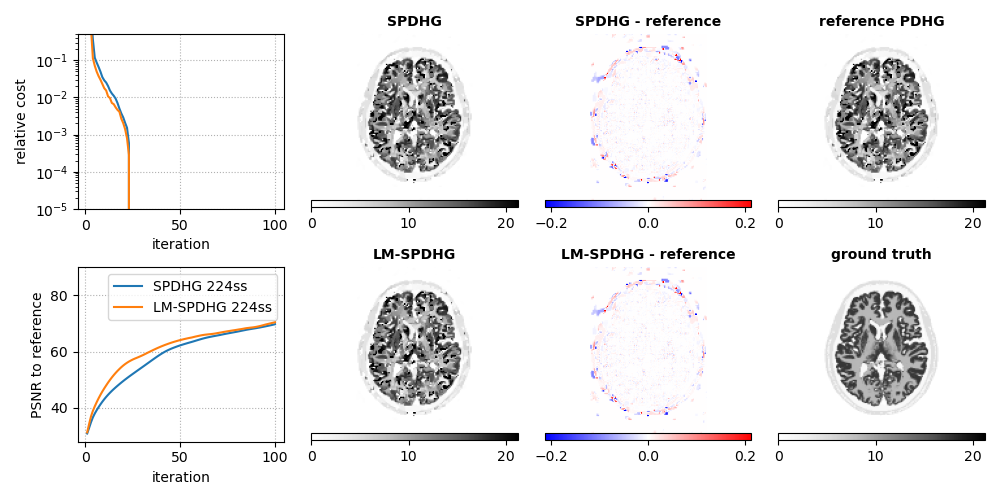}
    \caption{3e5 true (5e5 prompt) counts, DTV prior, $\beta = 0.1$}
  \end{subfigure}
  \caption{Comparison of convergence of sinogram SPDHG and LM-SPDHG
           for using 244 subsets.
           All three subfigures (a) - (c) show the same quantities but for
           different count levels and priors.
           Left column: relative cost and PSNR to with respect to the reference solution as
           a function of the iterations for sinogram SPDHG (blue) and LM-SPDHG (orange).
           Second column: SPDHG (top) and LM-SPDHG (bottom) reconstruction after 100 iterations.
           Third column: absolute difference between SPDHG/LM-SPDHG and reference PDHG reconstruction.
           Forth column: reference PDHG reconstruction using 20000 iterations (top) and ground truth
           used to generate the data (bottom).}
  \label{fig:lm-spdhg-var}
\end{figure*}

Figure~\ref{fig:lm-spdhg-var} summarizes the convergence comparison between sinogram SPDHG and 
LM-SPDHG for the same data set and prior as described in the previous subsection using the 
same warm start for both algorithms.
As demonstrated by the convergence metrics and the reconstructed images, the convergence of
sinogram SPDHG and LM-SPDHG is almost identical.
This also holds for different count levels and for the structural DTV priors 
as shown in subfigures (b) and (c) and for different levels of regularization as shown
in supplementary Figs.~1 and 2.
For the example with the DTV prior shown in subfigure (c), the convergence in terms of
PSNR seems to be even slightly faster with LM-SPDHG, which can be also
seen in the difference images with respect to the reference reconstruction.

\subsection*{Convergence of LM-SPDHG compared to listmode EM-TV}

\begin{figure*}
  \centering
  \begin{subfigure}[]{1.0\textwidth}
    \centering
    \includegraphics[width=0.8\textwidth]{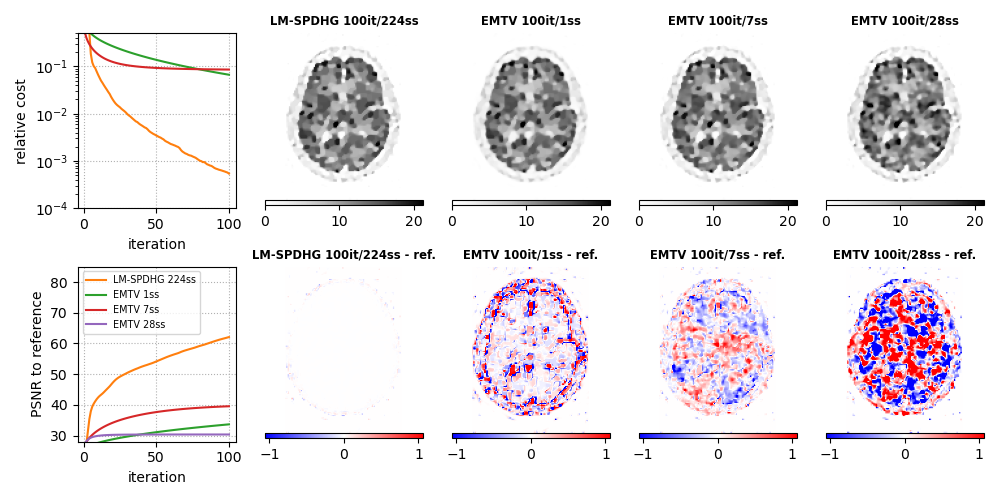}
    \caption{3e5 true (5e5 prompt) counts, TV prior, $\beta = 0.03$}
  \end{subfigure}
  \vfill
  \begin{subfigure}[]{1.0\textwidth}
    \centering
    \includegraphics[width=0.8\textwidth]{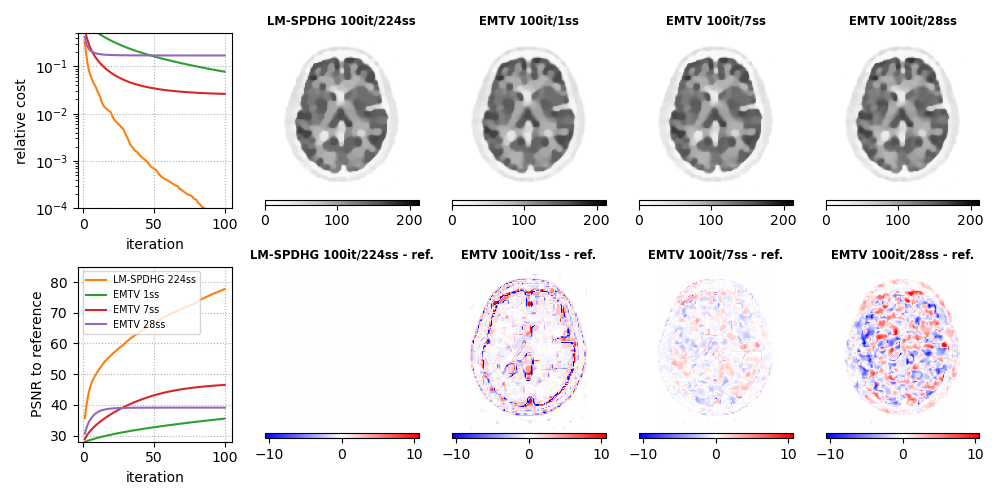}
    \caption{3e6 true (5e6 prompt) counts, TV prior, $\beta = 0.03$}
  \end{subfigure}
  \vfill
  \begin{subfigure}[]{1.0\textwidth}
    \centering
    \includegraphics[width=0.8\textwidth]{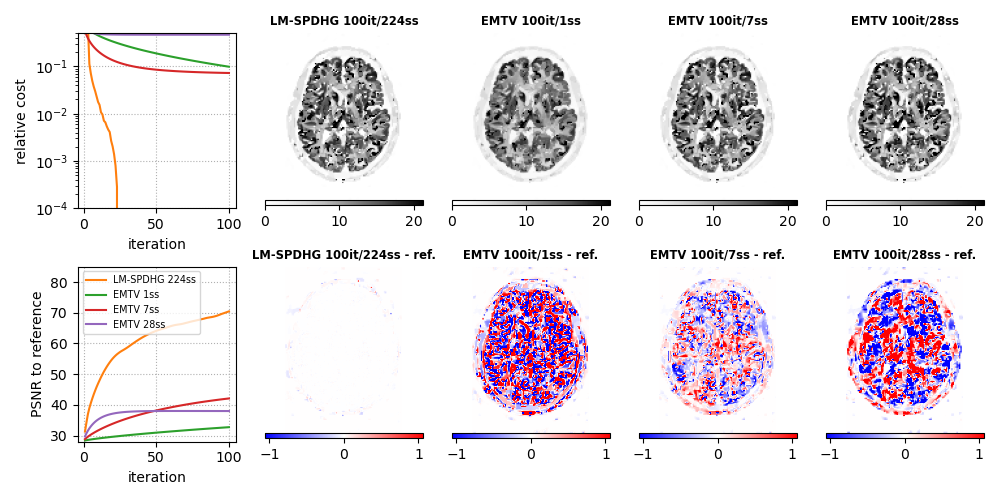}
    \caption{3e5 true (5e5 prompt) counts, DTV prior, $\beta = 0.1$}
  \end{subfigure}
  \caption{Same as Fig.~\ref{fig:lm-spdhg-var} but comparing the convergence of LM-SPDHG using
           224 subsets and listmode EM-TV using 1, 7 and 28 subsets.
           The top row of images shows the reconstruction after 100 iterations and the bottom
           row shows the difference to the reference reconstruction which is shown in
           Fig.~\ref{fig:lm-spdhg-var}.
           Note that the width of the color window in the difference plots is five times wider
           compared to Fig.~\ref{fig:lm-spdhg-var}.}
  \label{fig:emtv}
\end{figure*}

A comparison between the convergence of LM-SPDHG using 224 subsets and listmode 
EM-TV using 1, 7 and 28 subsets is shown in Fig.~\ref{fig:emtv}.
In all sub figures, LM-SPDHG converges much faster than EM-TV.
Interestingly, when using listmode EM-TV with more than one subset, the convergence
metrics saturate meaning that the algorithm \replaced{remains}{probably is stuck}
on a limited cycle and the optimal solution is not reached.
When using one subset, the convergence of EM-TV is much slower but does not seem
to saturate after 100 iterations.

\subsection*{Convergence vs number of subsets}

\begin{figure*}
  \centering
  \begin{subfigure}[b]{0.23\textwidth}
    \centering
    \includegraphics[width=1.0\textwidth]{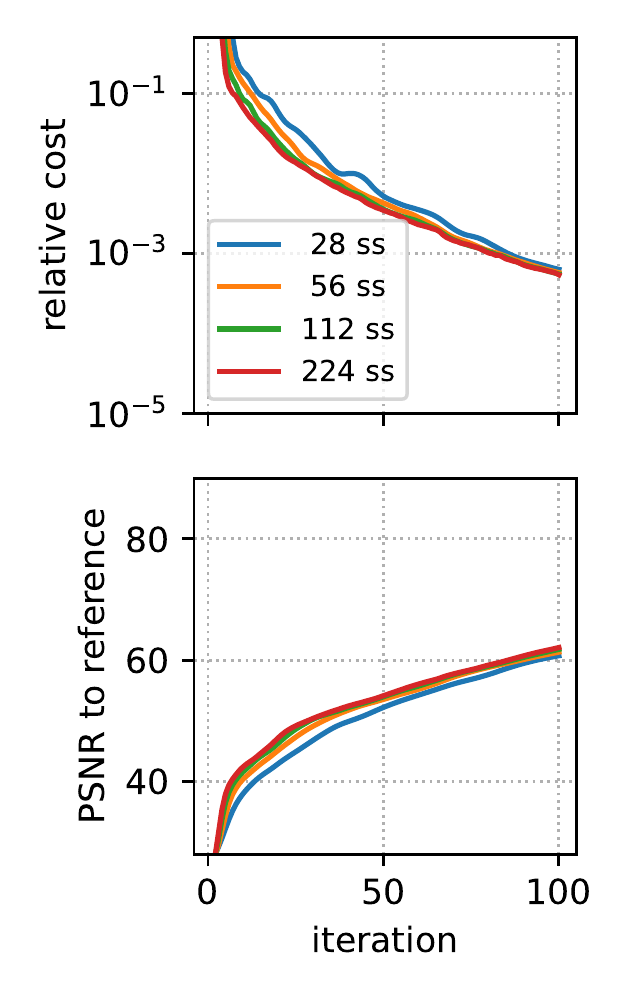}
    \caption{3e5 true (5e5 prompt) counts, TV prior, $\beta = 0.03$}
  \end{subfigure}
  \hfill
  \begin{subfigure}[b]{0.23\textwidth}
    \centering
    \includegraphics[width=1.0\textwidth]{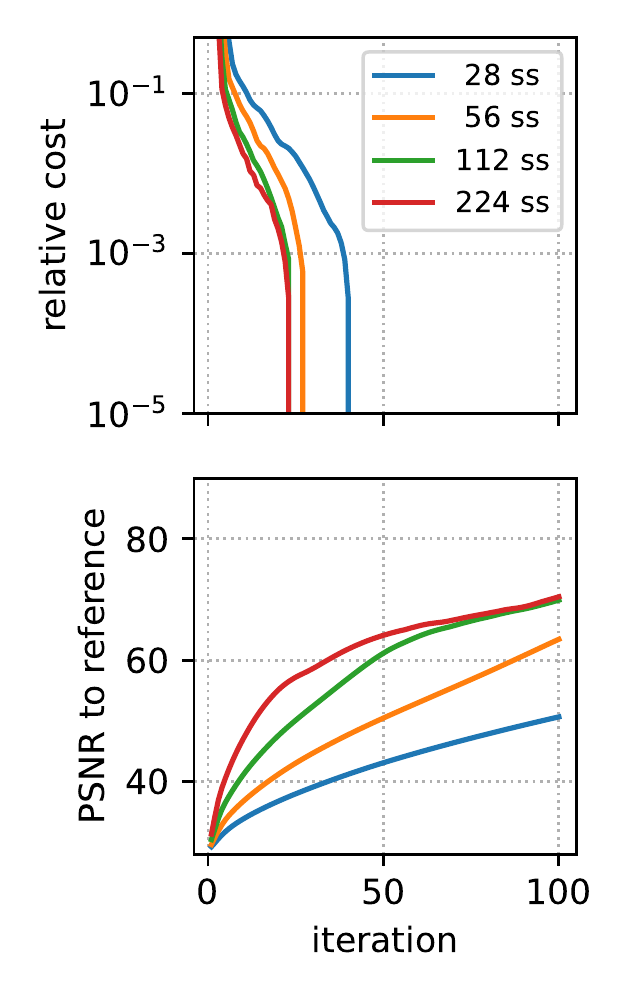}
    \caption{3e5 true (5e5 prompt) counts, DTV prior, $\beta = 0.1$}
  \end{subfigure}
  \hfill
  \begin{subfigure}[b]{0.23\textwidth}
    \centering
    \includegraphics[width=1.0\textwidth]{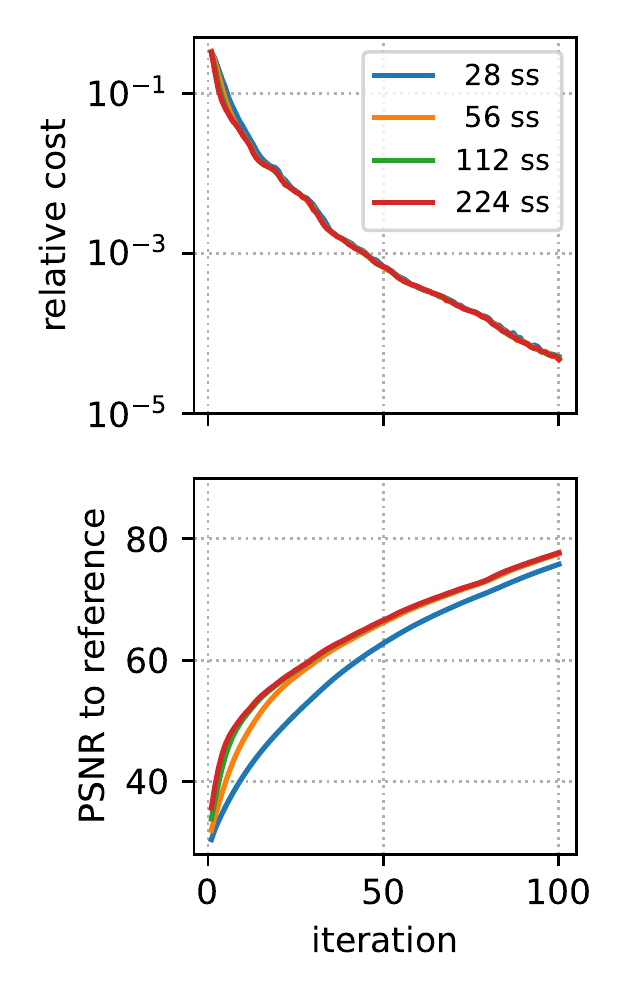}
    \caption{3e6 true (5e5 prompt) counts, TV prior, $\beta = 0.03$}
  \end{subfigure}
  \hfill
  \begin{subfigure}[b]{0.23\textwidth}
    \centering
    \includegraphics[width=1.0\textwidth]{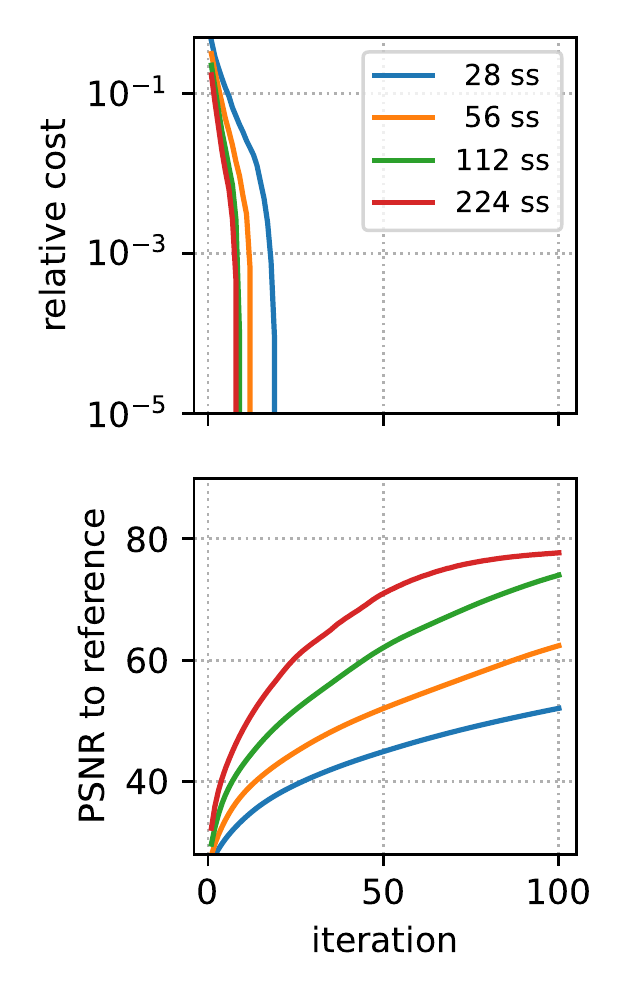}
    \caption{3e6 true (5e5 prompt) counts, DTV prior, $\beta = 0.1$}
  \end{subfigure}

  \caption{Converge of LM-SPDHG for different number of subsets at two count levels for the TV and 
           DTV prior. Note that when increasing the number of data subsets, the number of gradient
           updates per iteration increases as well.}
  \label{fig:num_subsets}
\end{figure*}

Figure~\ref{fig:num_subsets} shows the convergence of LM-SPDHG as a function of the number of data
subsets used for low (3e5) and high (3e6) counts and the TV and DTV prior.
In general, using 224 subsets leads to faster convergence compared to 56 and 112 subsets.
However, for the TV prior, especially at low counts, there is almost no difference between using
56, 112, and 224 subsets.
In contrast, for the DTV prior, the difference between using 56, 112, and 224 subsets is more
pronounced.
For a fixed number of subsets, convergence is faster for high count compared to low count data sets.

\subsection*{Reconstrution of 3D XCAT data}

\begin{figure*}
  \centering
    \includegraphics[width=1.0\textwidth]{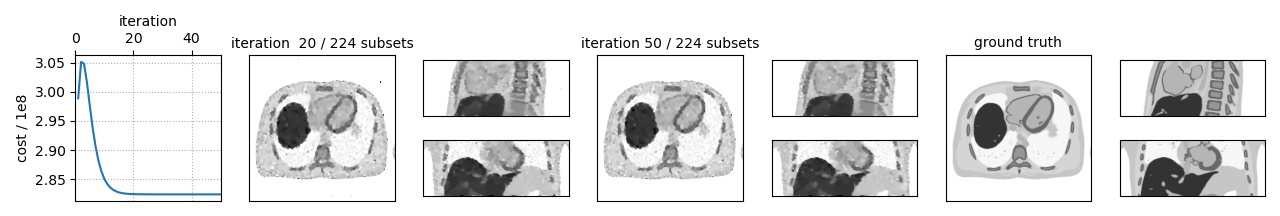}
  \caption{Convergence of LM-SPDHG for reconstruction of 3D TOF data generated from the XCAT phantom
           using 4e7 true (7e7) prompt counts and a TV prior with $\beta = 0.03$.
           Left column: evolution of cost function. 
           Columns 2-3: transversal, sagittal and coronal
           slice of LM-SPDHG reconstruction after 20 iterations and 224 subsets.
           Columns 4-5: LM-SPDHG reconstruction after 100 iterations and 224 subsets.
           Columns 6-7: Ground truth image.}
  \label{fig:xcat}
\end{figure*}

Figure~\ref{fig:xcat} shows the results of the reconstruction of the listmode data simulated
from the 3D XCAT phantom using a TV prior with $\beta = 0.03$. 
From the plot of the cost function and the reconstructions shown in the middle, we can see
that reasonable convergence is reached after approximately 25 iterations with 224 subsets.
Note that in contrast to our 2D experiments, the cost function initially increases until the 3rd iteration before it decreases and stabilizes after around 25 iterations.
As mentioned above, calculating the projections in listmode for 7e7 counts is roughly a factor
of 3 faster compared to sinogram-based processing which substantially speeds up every iteration.
Note that in our proof-of-concept implementation, the effective speed up is less than a 
factor of 3 since we did not implement the gradient operators and proximal mappings on a GPU yet
which leads to overhead due to gradient-based updates.
However, we expect that, once properly implemented on GPUs, this overhead will be small compared to
the time needed to calculate the actual TOF projections.

\subsection*{Reconstrution of NEMQ IQ phantom data}

\begin{figure*}
  \centering
    \includegraphics[width=1.0\textwidth]{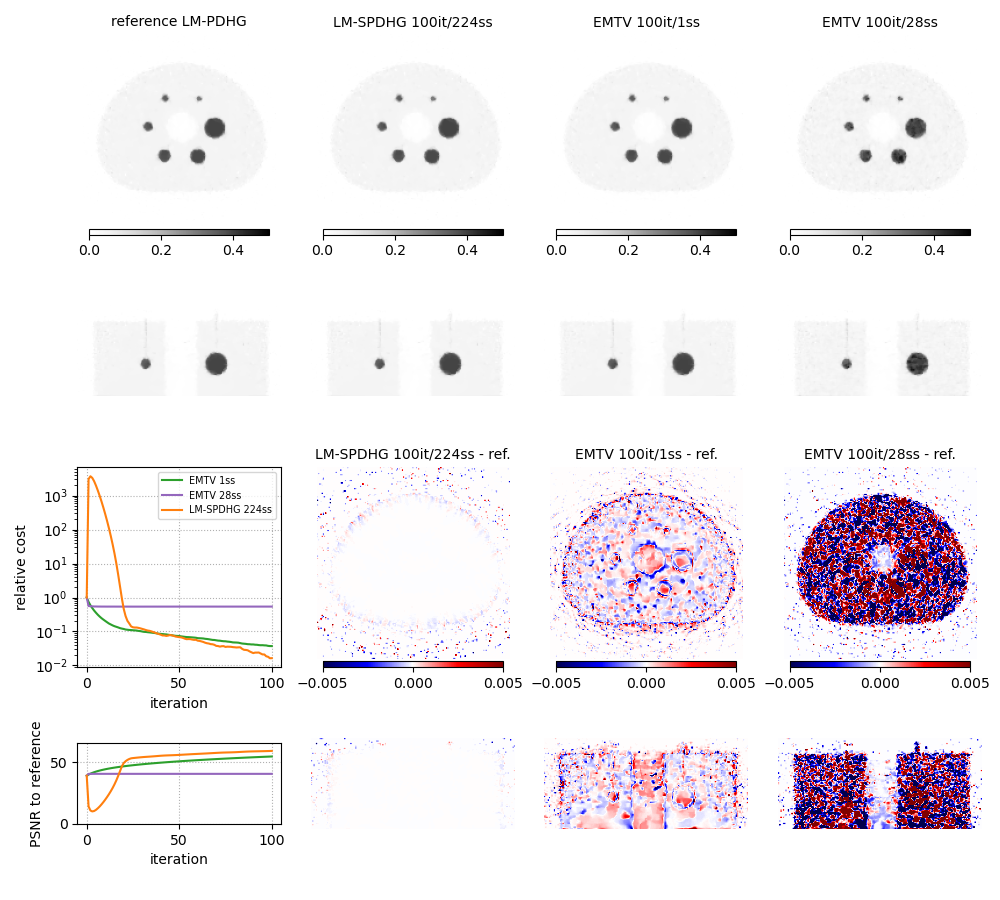}
  \caption{\added{
           Convergence of LM-SPDHG for reconstruction of real 3D TOF data from an acqusition of
           the NEMA image quality phantom on a GE Discovey MI 4 ring TOF PET system with a TV prior
           and $\beta = 6$ with 2.2e7 true (4.7e7 prompt) counts.
           First column, top two rows: transaxial and coronal slice of reference LM-PDHG 
           reconstruction using 20000 iterations and 1 subset. 
           Second column, top two rows: LM-SPDHG reconstruction using 100 iterations and
           224 subsets. 
           Third column, top two rows: EM-TV reconstruction using 100 iterations and
           1 subset. 
           Fourth column, top two rows: EM-TV reconstruction using 100 iterations and
           28 subsets. 
           Fist column, bottom two rows: relative cost and PSNR to reference reconstruction.
           Second, third, fourth column, bottom two rows: absolute difference compared to
           reference reconstruction.
           }}
  \label{fig:nema}
\end{figure*}

\added{
Figure~\ref{fig:nema} shows the results of the reconstructions of the listmode data from the 
acquisition of the NEMA image quality phantom on the GE Discovey MI 4 ring TOF PET system.
After 100 iterations, LM-SPDHG using 224 subsets has lower cost compared to both EM-TV
reconstructions and is closest to the referene reconstruction in terms of PSNR.
From the differene images it can be seen that the stochastic LM-SDPHG mostly differs from the
reference in the background outside the phantom.
In contrast, both EM-TV reconstructions show bigger residual differences within the phantom 
which are most pronounced in the ``warm'' background for EM-TV using 28 subsets and in the 
``cold'' central insert for EM-TV using 1 subset.
}


\section{Discussion}

The results of our numerical experiments presented in this work demonstrate that the speed of 
convergence of LM-SPDHG is essentially the same as the one of the original SPDHG using
sinograms.
For clinical acquisitions with state-of-the-art TOF PET scanners resulting in extremely
sparse data, LM-SPDHG has two distinct advantages compared to SPDHG.
First, during the iterations all forward and back projections can be performed in listmode
which is faster compared to sinogram projectors for sparse TOF data.

Second, as shown in Table~\ref{tab:mem}, the memory requirements are substantially reduced such
that even for an aquisition with 5e8 prompt coincidences only 12.5\,GB of memory are needed.
This actually enables a pure GPU-implementation of LM-SPDHG avoiding intermediate memory
transfer between the host and a state-of-the-art GPU with a memory of approximately 16\,GB.
Note that in our proof of concept implementation of LM-SPDHG we used a hybrid computing
approach where only the TOF PET forward and back projections where computed using a GPU.
We expect that once properly implemented purely on a GPU, the computation time required
for LM-SPDHG will further decrease since memory transfter between host and GPU is a bottleneck 
in our implementation.
Note that a pure GPU implementation of the conventional SPDHG algorithm for modern TOF PET
data is more complicated since more than 50\,GB of GPU is required.

In this proof of concept work, we only used two non-smooth TV-like priors
to benchmark LM-SPDHG. 
Note that in general, SPDHG and LM-SPDHG can be also used for smooth priors as long
as \added{the} proximal operator of the convex dual of the prior\added{s} can be efficiently 
calculated which is
e.g. the case for prior penalizing the squared L2 norm or Huber TV.
However, for smooth priors the speed of convergence of (LM)-SPDHG should be benchmarked against
other stochastic gradient-based optimization techniques such as the stochastic variance reduced gradient (SVRG) \cite{Johnson2013} or SAGA \cite{Defazio2014} which is beyond the scope of this work.
We also note that in this work we used a scalar spatially-invariant prior strength $\beta$ which
leads to a spatially-variant local pertubation response (LPR). 
For pratical applications where a spatially-invariant reponse is desirable, a spatially-varient prior 
strength should be used as described in \cite{Ahn2008,Tsai2020}.
We emphasize that the listmode EM-TV algorithm
is still a very practical and useful algorithm to approximate the solution 
of the optimization problem \eqref{eq:primal} for listmode data and non-smooth priors.
However, as shown in Fig.~\ref{fig:emtv}, it seems that when using ordered subsets for acceleration,
EM-TV does not reach the optimal solution but rather remains on a limit cycle similar to the
behaviour of OSEM.
Whether the difference between the limit cycle and the true optimal solution is of importance
for a given count level and clinical task, should be investigated in the future.

\added{As shown in Fig.~\ref{fig:num_subsets}, choosing the ``optimal'' number of subsets 
for LM-SPDHG seems to depend on the prior and the number of acquired counts.
For a reconstruction of a high-count scan with DTV, it is possible that LM-SPDHG with more
than 224 subsets converges even faster.
A detailed investigation of the choice of the number of subsets for different acquistion
and reconstruction scenarios is left for future research.}

Last but not least, we would like to note that according to our experience, SPDHG and LM-SPDHG
for TOF PET reconstruction also converge for $\rho$ slightly greater than one which might
be of interest for practical applications where stopping as early as possible can be important.
Suppl. Fig.~3 shows that in our 3D XCAT example, convergence can be further accelerated by
using $\rho = 8$ and $\gamma = 30/ \|x^0\|_\infty$.
A detailed investigation of the convergence speed as a function of $\rho$ and $\gamma$ potentially
involving an iteration-dependent choice of $\rho$ \cite{Goldstein2015} is, however, 
beyond the scope of this manuscript and left for future research.

\section{Conclusion}

For sparse TOF data, the proposed LM-SPDHG algorithm severely reduces the memory requirements 
and computation times compared to the original SPDHG enabling the application of LM-SPDHG 
in routine clinical practice where short reconstruction times are crucial.

\section*{Acknowledgements}

The authors wold like to thank Claire Delplancke for the insightful discussion on the
initialization of (S)PDHG.
\added{Moreover, the authors would like to thank Prof. Kristof Baete for the acquisition
of the NEMA data set and Dr. Ahmadreza Rezaei for preprocessing of the same data set.}
This work was supported in part by the NIH grant 1P41EB017183-01A1.

\begin{appendices}

\section{Listmode re-formulation of the sinogram-based minimization problem}
\label{appendix:lm}
\added{
Here, as basis for the proposed LM-SPDHG algorithm, we provide details on the listmode re-formulation 
of the sinogram-based minimization problem \eqref{eq:primal}. To this aim, we denote the dependence 
of $D_i(\dot)$ on the data $d_i$ explicitly via $D(\cdot; d_i)$. Further, recall that for $N$ the list 
of detected events, $i_e$ is the sinogram bin in which event $e \in N$ was detected. 
With $e_0 \notin N$ denoting an additional, artificial event summarizing zero measurements we then obtain
\begin{align*}
\sum_{i=1}^m D((Px + s)_i; d_i)
& = \sum_{i=1}^m \sum_{e \in N: i_e = i} \frac{(Px + s)_{i_e}}{d_{i_e}} - \log \left((Px+ s)_{i_e} \right) + \sum_{ i \,  : \,   d_i = 0 } (Px)_i  \\
 & =  \sum_{e \in N} \frac{(Px+s)_{i_e}}{d_{i_e}}  - \log \left(\frac{(Px+s)_{i_e}}{d_{i_e}} \right)
 -  \sum_{e \in N} \log (d_{i_e}) +  \sum_{ i \,  : \,   d_i = 0 } (Px)_i  \\ 
& =  \sum_{e \in N \cup \{e_0\}} D \left( \frac{(P^{LM}_N x+s^{LM})_{e}}{d_{i_e}};  d^{LM}_e \right) -  \sum_{e \in N} \log (d_{i_e}) \\ 
& =  \sum_{e \in N \cup \{e_0\}} D \left(\hat P^{LM}_N x+\hat s^{LM})_{e} ; d^{LM}_e \right)  - \sum_{e \in N} \log (d_{i_e}) \\ 
\end{align*}
where $d^{LM}_e=1$ for $e \in N$, $d_{e_0}^{LM} = 0$, $s^{LM}$ with $s^{LM}_e = s_{i_e}$ is the list-mode random and scatter estimate and
\[ (P^{LM} u) _e = \begin{cases}
 (Pu)_{i_e} & \text{if } e \in N \\
 \sum_{i \,  : \,   d_i = 0} (Pu)_{i} & \text{if $e = {e_0}$}.
\end{cases}
\]
The last step uses the rescaled operator $ \hat P^{LM} $ and the rescaled estimate  $\hat s^{LM}$ defined as $ (\hat P^{LM} u)_e= (P^{LM} u)_e/d_{i_e}$ and $(\hat s^{LM})_e = s^{LM}/ d_{i_e}$ for $e \in N$, and $ (\hat P^{LM} u)_{e_0} = (P^{LM} u)_{e_0}$ and $(\hat s^{LM})_{e_0} = 0$, respectively.
}

\added{
Now the last line of the above reformulation provides an equivalent data term which (after removing 
the constant $\sum_{e \in N} \log (d_{i_e})$) can be used to obtain an equivalent list-mode-based 
reformulation of \eqref{eq:primal}.  
Applying the SPDHG for this reformulation results in the proposed LM-SPDHG Algorithm \ref{alg:lmspdhg}. 
Note that there, the non-rescaled operator $P^{LM}$ and estimate $s^{LM}$ appear since i) on line 11 
of the algorithm, the rescaling of $P^{LM}$ and $s^{LM}$ cancels out with the rescaling of $P^{LM}$ 
in the step size $S_k$ of \eqref{eq:lm_stepsizes}, and ii) the rescaling of $P^{LM}$ in line 12 is 
incorporated explicitly. 
Furthermore, note that we use the full forward operator $P$ instead of $P^{LM}_{N_k}$ for the stepsize 
$T_k$, for reasons explained in the paragraph above the respective equation \eqref{eq:lm_stepsizes}.
}
\end{appendices}

\printbibliography

\end{document}


\renewcommand{\figurename}{Supplementary Figure}

\begin{figure*}
  \centering
  \begin{subfigure}[]{1.0\textwidth}
    \centering
    \includegraphics[width=0.9\textwidth]{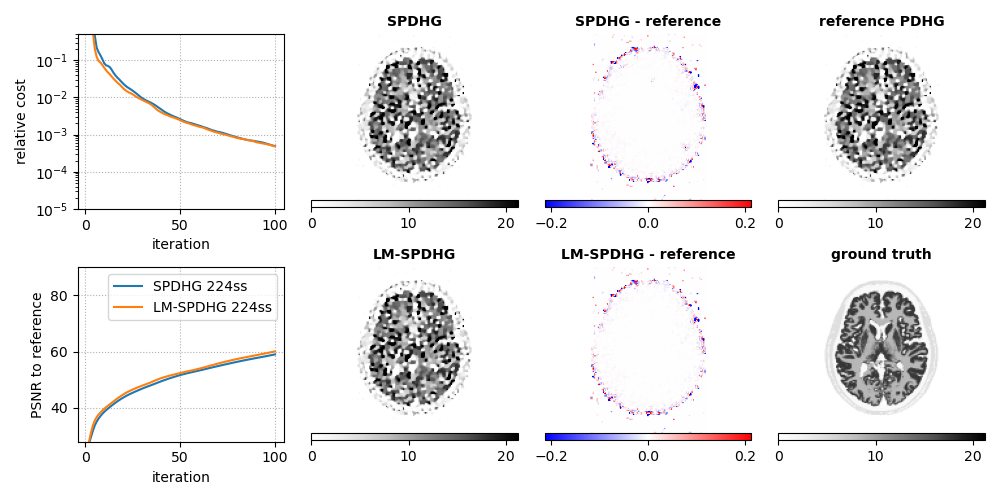}
    \caption{3e5 true (5e5 prompt) counts, TV prior, $\beta = 0.01$}
  \end{subfigure}
  \vfill
  \begin{subfigure}[]{1.0\textwidth}
    \centering
    \includegraphics[width=0.9\textwidth]{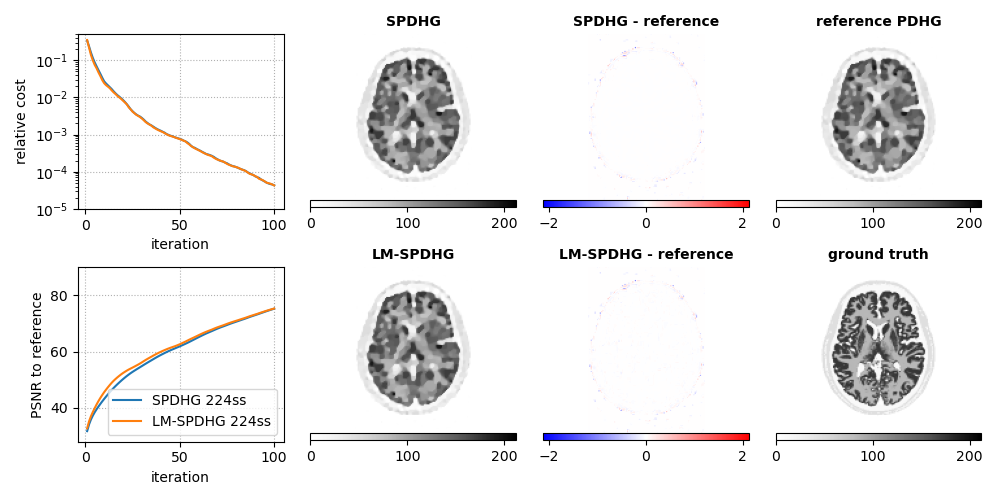}
    \caption{3e6 true (5e6 prompt) counts, TV prior, $\beta = 0.01$}
  \end{subfigure}
  \caption{Same as Fig.~3 for a weaker TV prior for low (top) and high counts (bottom).}
\end{figure*}

\begin{figure*}
  \centering
  \begin{subfigure}[]{1.0\textwidth}
    \centering
    \includegraphics[width=0.9\textwidth]{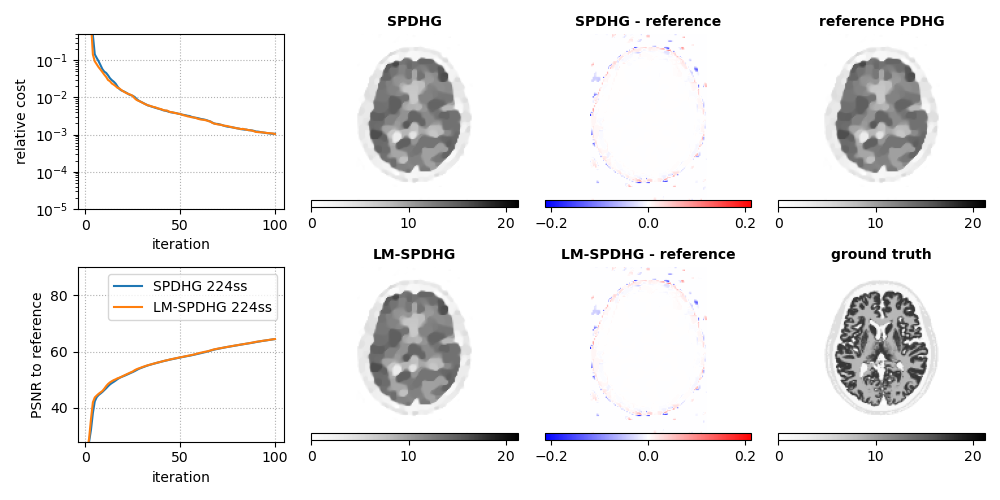}
    \caption{3e5 true (5e5 prompt) counts, TV prior, $\beta = 0.1$}
  \end{subfigure}
  \vfill
  \begin{subfigure}[]{1.0\textwidth}
    \centering
    \includegraphics[width=0.9\textwidth]{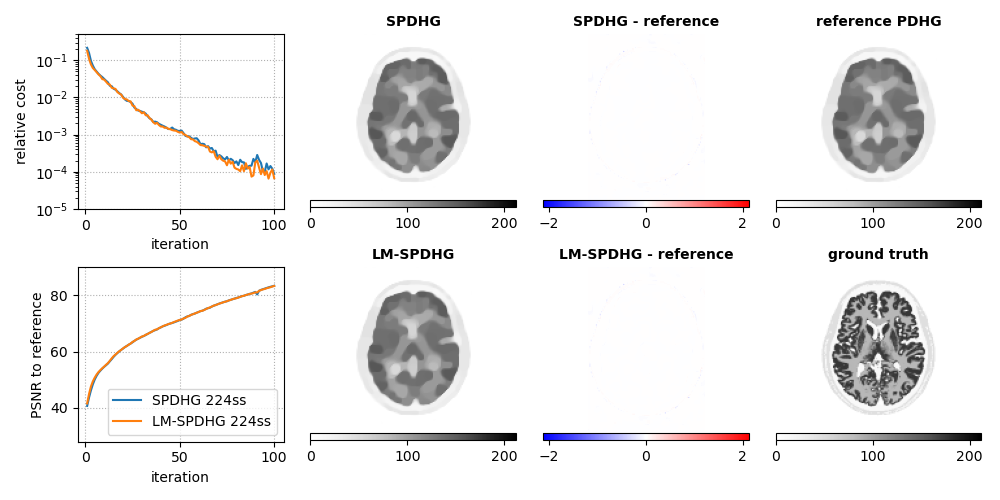}
    \caption{3e6 true (5e6 prompt) counts, TV prior, $\beta = 0.1$}
  \end{subfigure}
  \caption{Same as Fig.~3 for a stronger TV prior for low (top) and high counts (bottom).}
\end{figure*}

\begin{figure*}
  \centering
  \begin{subfigure}[]{1.0\textwidth}
    \centering
    \includegraphics[width=0.6\textwidth]{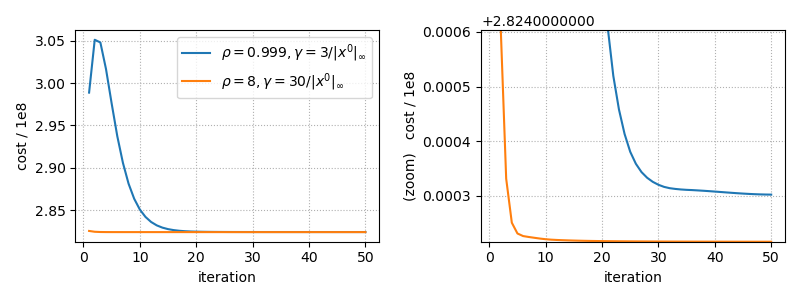}
    \caption{Evolution of the cost function for two choices of $\rho$ and $\gamma$ for the 3D
             XCAT data set.}
  \end{subfigure}
  \vfill
  \begin{subfigure}[]{1.0\textwidth}
    \centering
    \includegraphics[width=1.0\textwidth]{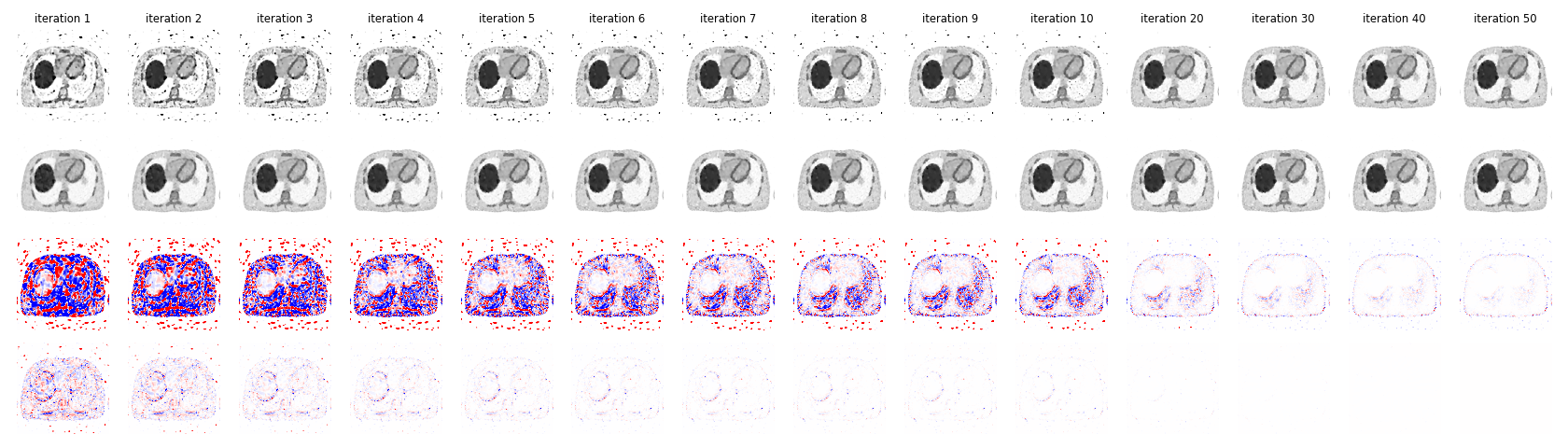}
    \caption{row 1/2: intermediate reconstructions at different iteration numbers for
                      $\rho = 0.999, \ \gamma = 3/ \| x^0 \|_\infty$ (row 1) and
                      $\rho = 8, \ \gamma = 30/ \| x^0 \|_\infty$ (row2). 
             row 3/4: absolute difference compared to the reconstruction at iteration 50
                      using ($\rho = 8, \ \gamma = 30/ \| x^0 \|_\infty$).
                      The minimum / maximum of the red-white-blue colormap is -5\% / +5\%
                      of the maximum of the reconstruction at iteration 50.}
  \end{subfigure}
  \caption{Same as Fig.~6 but including a comparison with LM-SPDHG using $\rho = 8$ and 
           $\gamma = 30 / \| x^0 \|_\infty$. 
           The latter substantially increases the speed of convergence.
           Note that increasing $\gamma$ and keeping $\rho = 0.999$ did not increase the
           speed of convergence.}
\end{figure*}